\documentclass[11pt,a4paper]{article}
\pdfoutput=1

\usepackage{jheppub}
\usepackage{rotating}

\usepackage{wrapfig}
\usepackage{amsfonts}
\usepackage{amsmath}
\usepackage{slashed}
\usepackage{amssymb}
\usepackage{latexsym}
\usepackage{multirow}
\usepackage{hyperref}
\usepackage{enumitem}
\hypersetup{colorlinks=true,citecolor=red,linkcolor=magenta,urlcolor=blue}
\usepackage{subfigure}
\usepackage{relsize}
\usepackage{booktabs}
\usepackage{cleveref}
\usepackage{fancyhdr}
\usepackage{float}
\usepackage{graphicx}
\graphicspath{ {./Figures/} }
\usepackage{microtype}
\usepackage{tabularx}
\usepackage{xcolor}
\usepackage{orcidlink}
\usepackage[bottom]{footmisc}

\title{Resurrecting the LHC discovery potential in the extended Type-II Seesaw Model}
\abstract{The juxtaposition of the precision of lepton flavour measurements and the limited energy range of the Large Hadron Collider (LHC) to discover dynamical degrees of freedom linked to the generation of the observed lepton mass patterns naively suggests only a limited relevance of the LHC's high luminosity phase. This, potentially, extends to future colliders. Using the concrete example of the type-II seesaw model and its effective field theory extension, we show that blind directions create a rich phenomenological interplay of muon precision measurements and electroweak resonance searches at present and future colliders, with testable implications for the HL-LHC phase.}
\author[a]{Upalaparna Banerjee\orcidlink{0000-0002-5222-1900},} 
\author[b]{Christoph Englert\orcidlink{0000-0003-2201-0667},} 
\author[b]{Wrishik Naskar\orcidlink{0000-0002-4357-8991}}
\affiliation[a]{Indian Institute of Technology Kanpur, Kalyanpur, Kanpur 208016, Uttar Pradesh, India}
\affiliation[b]{School of Physics \& Astronomy, University of Glasgow, Glasgow G12 8QQ, United Kingdom}

\emailAdd{upalab@iitk.ac.in} 
\emailAdd{christoph.englert@glasgow.ac.uk}
\emailAdd{w.naskar.1@research.gla.ac.uk}
\begin{document}
\maketitle
\flushbottom
\allowdisplaybreaks
	
\section{Introduction}
\label{sec:intro}
The observed mass hierarchies in the lepton sector of the Standard Model (SM) of Particle Physics present a motivated source for physics beyond the SM. The history of, e.g., dynamically explaining the smallness of neutrino masses~\cite{Kajita:2016cak,McDonald:2016ixn,KamLAND:2002uet,K2K:2002icj} is as illustrious as it is long. What singles out the neutrinos in comparison to the charged leptons is the plethora of model-building avenues that present themselves to generate mass terms. These can rely on the direct inclusion of right-handed neutrino SM singlets~\cite{Minkowski:1977sc,Mohapatra:1979ia}, on extending the scalar sector of the SM~\cite{Konetschny:1977bn,Magg:1980ut,Schechter:1980gr,Cheng:1980qt,Mohapatra:1980yp,Lazarides:1980nt}, or on employing direct couplings to matter to source mass terms via loop-effects~\cite{Zee:1985id,Babu:1988ki}. Given that the neutrinos are related to the charged sector via gauge-invariance of the SM, the precision measurements that can be performed in the charged lepton sector have a wider phenomenological relevance in these scenarios. In particular for models with extra scalars, the requirement of non-trivial electroweak quantum numbers makes these states excellent targets for exotics searches at colliders, e.g., through Drell-Yan-like production~as discussed in \cite{Han:2007bk,Anisha:2021jlz} (see Refs.~\cite{CMS:2017pet,ATLAS:2017xqs,ATLAS:2018ceg,ATLAS:2022pbd} for recent searches at the Large Hadron Collider, LHC). The charged scalars have a rich phenomenology and can also be probed through low-energy meson decays, through off-shell production~\cite{Bambhaniya:2015nea,Wang:2018bgp}.

It is conceivable that mass scale of the degrees that underpin the generation of neutrino masses lies outside the direct sensitivity range of the LHC. Whilst entirely possible, and even expected in, e.g., type-II seesaw mechanism scenarios, the relevant energy scales could even lie beyond the reach of future machines. However, if the new physics scenario does indeed deviate from its simplest implementation, new phenomenological implications can change this picture. In the context of seesaw mechanisms, in particular the precision measurements of $\mu \to 3e$ and $\mu \to e \gamma$ provide tight constraints that quickly push the mass scale of the new BSM degrees of freedom outside the LHC discovery potential. 

In this work, to explore a related but different avenue, we consider modifications of the type-II seesaw model by including higher dimensional effective operators. Identifying a relevant set of couplings, we show that consistency with precision electroweak and muon data could still be compatible with resonances in the LHC-accessible energy regime. This comes at the price of a TeV-scale departure from the standard type-II scenario, which provides a testable implication in its own right. This work is organized as follows. In Sec.~\ref{sec:model} we swiftly introduce the type-II seesaw scenario to make this work self-contained. We then consider constraints on this model class in Sec.~\ref{sec:constraints}. In Sec.~\ref{sec:eftdef}, we review these findings from the vantage point of non-minimal, TeV-scale modifications. These can move mass scales to a range that is probable at present and future colliders. We comment on the connection of the low-energy precision observables (we focus on muon physics) and its TeV scale implications via the renormalisation group flow in Sec.~\ref{subsec:rge} before concluding in Sec.~\ref{sec:conclusions}.

\section{Phenomenology of the (deformed) Type-II Seesaw Model}
\subsection{The model}
\label{sec:model}
The type-II seesaw offers a natural and a minimal framework to account for the smallness of observed neutrino masses, which comes about by extending the SM scalar sector minimally by an $\text{SU}(2)_L$ triplet scalar $\Delta$, characterized by a hypercharge $Y_\Delta = 1$. The inclusion of $\Delta$ leads to new terms in the scalar potential $V(\Phi, \Delta)$, 
\begin{equation}
\label{eqn:scalarpot}
    \begin{split}
        V(\Phi,\Delta) = -m_H^2 \,(\Phi^\dagger \Phi) + \lambda_H (\Phi^\dagger \Phi)^2 + & m_\Delta^2 \text{Tr} [\Delta^\dagger \Delta] + \lambda_{\Delta_1} (\text{Tr} [\Delta^\dagger \Delta])^2 + \lambda_{\Delta_2} \text{Tr} [(\Delta^\dagger \Delta)^2] \\
        + \lambda_{\Delta_3} (\Phi^\dagger \Phi)\text{Tr} [\Delta^\dagger \Delta] + & \lambda_{\Delta_4} \Phi^\dagger \Delta \Delta^\dagger \Phi +\left[ \mu_{\Delta} \Phi^\dagger i\sigma^2 \Delta^\dagger \Phi + \text{h.c.}\right]\;,
    \end{split}
\end{equation}
where $\Phi$ denotes the original SM $\text{SU}(2)_L$ doublet scalar. In addition to the neutral components, $\Delta$ also comprises singly-charged and doubly-charged scalars, holding significant phenomenological implications as direct probes of the type-II seesaw mechanism and its characteristics in various collider experiments~\cite{Chakrabortty:2015zpm,Primulando:2019evb,Fuks:2019clu,Cai:2017mow,Antusch:2018svb,BhupalDev:2018tox,delAguila:2013mia,ATLAS:2017xqs,CMS:2022cbe}. We can explicitly express $\Phi$ and $\Delta$ in the following manner by expanding around their vacuum expectation values (vevs) denoted by $v_\Phi$ and $v_\Delta$, respectively,
\begin{equation}\label{eq:scalar-expression}
	\Phi = {1\over \sqrt{2}}\begin{pmatrix}
		\sqrt{2}\phi^+ \\
		\,(\phi+v_\Phi+i\eta)
	\end{pmatrix},
	\quad \Delta = {1\over \sqrt{2}} \begin{pmatrix}
		{\delta^+} &  {\sqrt{2}}\Delta^{++} \\	
		(\delta^0+v_\Delta+i\chi) & -\delta^+
	\end{pmatrix}\;.
\end{equation}
Most importantly, the addition of the complex triplet results in novel Yukawa interactions between $\Delta$ and the left-handed lepton doublet ($\psi_L$). These are quintessential for generating non-zero neutrino masses through the non-zero vev of the complex triplet $\Delta$ after electroweak symmetry breaking (EWSB),
\begin{equation}
\label{eqn:yuk}
    \mathcal{L^{\text{BSM}}_{\text{Yukawa}}} = - (Y_\Delta)_{ij} \Bar{\psi}^c_{L_i} \Delta \psi_{L_j} + \text{h.c.}\;.
\end{equation}
Here, $Y_\Delta$ is the BSM Yukawa coupling matrix. The Lagrangian for the type-II seesaw mechanism is therefore given by,
\begin{equation}
\mathcal{L^{\text{type-II}}} = \mathcal{L}_{\text{SM}} + \text{Tr}[D_\mu \Delta^\dagger D^\mu \Delta] - V(\Phi,\Delta) + \mathcal{L}^{\text{BSM}}_{\text{Yukawa}}\;.
\end{equation}
After EWSB, the two vevs $v_\Phi$ and $v_\Delta$ contribute to the masses of the gauge bosons, with, 
\begin{equation}\label{eq:gauge-boson-masses}
M^2_{W^\pm}={g^2\over 4} (v_\Phi^2+2 v_\Delta^2), \quad M^2_Z={g^2+g'^2\over 4}(v_\Phi^2+4 v_\Delta^2)\,,
\end{equation}
where $g$ and $g'$ are the coupling constants for the $SU(2)_L$ and $U(1)_{Y}$ gauge groups, respectively. The masses of the scalars become,
\begin{align}
	M_{\text{neutral,\,CP-even}}^2 &= \begin{pmatrix}
		2 \lambda_H v_{\Phi}^2 & -\sqrt{2}\mu_{\Delta}v_{\Phi}+v_{\Delta}v_{\Phi}(\lambda_{\Delta_3}+\lambda_{\Delta_4}) \\
		-\sqrt{2}\mu_{\Delta}v_{\Phi}+v_{\Delta}v_{\Phi}(\lambda_{\Delta_3}+\lambda_{\Delta_4}) & \cfrac{\mu_{\Delta}v_{\Phi}^2}{\sqrt{2}v_{\Delta}}+2v_{\Delta}^2(\lambda_{\Delta_1}+\lambda_{\Delta_2})
	\end{pmatrix}\;,\label{eq:neutral-mass-matrix-1}\\
	M_{\text{neutral,\,CP-odd}}^2 &= \begin{pmatrix}
		2 \sqrt{2} \mu_{\Delta} v_{\Delta} & -\sqrt{2}\mu_{\Delta}v_{\Phi} \\
		-\sqrt{2}\mu_{\Delta}v_{\Phi} & \cfrac{\mu_{\Delta}v_{\Phi}^2}{\sqrt{2}v_{\Delta}}
	\end{pmatrix}\;,\label{eq:neutral-mass-matrix-2}\\
	M_{\text{charged}}^2 &= \begin{pmatrix}
		\cfrac{\mu_{\Delta}v_{\Phi}^2}{\sqrt{2}v_{\Delta}}-\cfrac{1}{4}v_{\Phi}^2\lambda_{\Delta_4} & -\mu_{\Delta}v_{\Phi}+\cfrac{v_{\Phi}v_{\Delta}\lambda_{\Delta_4}}{2\sqrt{2}} \\
		-\mu_{\Delta}v_{\Phi}+\cfrac{v_{\Phi}v_{\Delta}\lambda_{\Delta_4}}{2\sqrt{2}} & \sqrt{2}\mu_{\Delta}v_{\Delta}-\cfrac{1}{2}v_{\Delta}^2\lambda_{\Delta_4}
	\end{pmatrix}\;,\label{eq:charged-mass-matrix}\\
	M_{\Delta^{\pm\pm}}^2 &= \cfrac{\mu_{\Delta}v_{\Phi}^2}{\sqrt{2}\,v_{\Delta}}-v_{\Delta}^2\lambda_{\Delta_2}-\cfrac{1}{2}v_{\Phi}^2\lambda_{\Delta_4}\;.
\end{align}
Since $\Delta$ acquires non-zero vev, the scalar sector, apart from the doubly charged scalar ($\Delta^{\pm\pm}$) involves non-trivial mixing, as can be seen in Eqs.~\eqref{eq:neutral-mass-matrix-1},~\eqref{eq:neutral-mass-matrix-2} and~\eqref{eq:charged-mass-matrix}. The physical states can be recovered in the following manner,
\begin{gather}
	\begin{pmatrix}
		h \\
		\Delta^0
	\end{pmatrix} = \begin{pmatrix}
		\cos{\alpha} & \sin{\alpha} \\
		-\sin{\alpha} & \cos{\alpha}
	\end{pmatrix}\begin{pmatrix}
		\phi \\
		\delta^0
	\end{pmatrix},\quad 
	\begin{pmatrix}
		G^0 \\
		A
	\end{pmatrix} = \begin{pmatrix}
		\cos{\beta_0} & \sin{\beta_0} \\
		-\sin{\beta_0} & \cos{\beta_0}
	\end{pmatrix}\begin{pmatrix}
		\eta \\
		\chi
	\end{pmatrix},\nonumber\\	\begin{pmatrix}
		G^{\pm} \\
		\Delta^{\pm}
	\end{pmatrix} = \begin{pmatrix}
		\cos{\beta_{\pm}} & \sin{\beta_{\pm}} \\
		-\sin{\beta_{\pm}} & \cos{\beta_{\pm}}
	\end{pmatrix}\begin{pmatrix}
		\phi^{\pm} \\
		\delta^{\pm}
	\end{pmatrix}\;,
\end{gather}
with
\begin{gather}
	\tan{2\alpha} = \frac{v_\Phi\left[\sqrt{2}\mu_{\Delta}-v_{\Delta}(\lambda_{\Delta_3}+\lambda_{\Delta_4})\right]}{\left[v_{\Phi}^2\left(\cfrac{1}{2\sqrt{2}}\cfrac{\mu_{\Delta}}{v_{\Delta}}-\lambda_H\right)+v_{\Delta}^2(\lambda_{\Delta_1}+\lambda_{\Delta_2})\right]}\,,\\
	\tan{\beta_0} = 2v_{\Delta}/v_{\Phi}\,,\qquad \tan{\beta_{\pm}} = \sqrt{2}v_{\Delta}/v_{\Phi}\,.
\end{gather}
In addition to the three Goldstone bosons ($G^{0}$ and $G^{\pm}$), this gives rise to additional singly-charged scalars $\Delta^{\pm}$, one CP-odd scalar $A$, and CP-even scalars $\Delta^{0}$ and $h$. We identify the latter with the observed $125$ GeV Higgs boson.  By diagonalising $M_{\text{neutral,CP-even}}^2$, $M_{\text{neutral,CP-odd}}^2$ and $M_{\text{charged}}^2$, we can determine the masses of the CP-even neutral scalars, CP-odd scalars and charged scalars respectively, while the masses of the Goldstone bosons are identically zero. The electroweak vev is defined as $v^2 \equiv v_\Phi^2 + 2 v_\Delta^2 \approx (246~\text{GeV})^2$, signalising a violation of custodial isospin: the ratio of the gauge boson masses is parameterised by the $\rho$-parameter, given by $\rho = M_W^2/(M_Z^2 \cos^2 \theta_W)$, and the current electroweak precision constraints on $\rho$, sets an upper limit on the triplet vev $v_\Delta < 4.8~\text{GeV}$~\cite{Antusch:2018svb,Primulando:2019evb} at 95\% CL. Upon EWSB, the new physics contributions to the Yukawa interactions in the neutrino sector are given by,
\begin{equation}
\mathcal{L^{\text{BSM}}_{\text{Yukawa}}} \supset \frac{v_{\Delta}}{ \sqrt{2}}\left[ (Y_\Delta + Y_\Delta^T)_{ij} \,\Bar{\nu}^{c}_i \nu_j \right]\;,
\end{equation}
where $i$, and $j$ are flavour indices. The neutrino mass-mixing matrix ($M_\nu$) originating from the Lagrangian is diagonalised by the unitary Pontecorvo--Maki--Nakagawa--Sakata (PMNS) matrix, 
\begin{equation}
M_\nu = U_{\text{PMNS}}^*\; M_{\nu}^{\text{diag}} \;U_{\text{PMNS}}^\dagger\;,
\end{equation}
where $M_{\nu}^{\text{diag}} = \text{diag}(m_{\nu_1}$, $m_{\nu_2}$, $m_{\nu_3})$. The different $m_{\nu_i}$ represent the three physical neutrino masses; the Yukawa matrix therefore is given by,
\begin{equation}
Y_\Delta = \frac{M_\nu}{\sqrt{2}v_\Delta}\;.
\end{equation}
The nature of the Yukawa matrix is thus determined by neutrino oscillation parameters and $v_\Delta$. Non-zero neutrino mass-splittings allow for non-zero flavour off-diagonal couplings, which can lead to interesting lepton flavour violating signatures~\cite{Dinh:2012bp,Chakrabortty:2012vp,Barrie:2022ake}, which we further discuss later in Sec.~\ref{subsubsec:lfv}.
\subsection{Constraints}
\label{sec:constraints}
The addition of the complex $SU(2)_L$ triplet gives rise to a large range of theoretical and phenomenological implications. The addition of several physical BSM scalars offer unique collider signatures, precision constraints, and novel interactions sensitive to various low-energy processes and unique decay channels. All of these approaches are effective in probing and constraining parts of the viable parameter space of the model. Constraints on this vast parameter space have been analysed in the existing type-II seesaw literature. In particular, Ref.~\cite{Primulando:2019evb} has provided combined theoretical constraints from vacuum stability, perturbative unitarity, electroweak precision and Higgs boson data in terms of the triplet mass splittings ($\Delta M = M_{\Delta^{\pm}} - M_{\Delta^{\pm \pm}}$) and $M_{\Delta^{\pm \pm}}$ recently, with implications for the LHC. With the LHC Run-3 in progress, its high-luminosity phase on the horizon, and future facilities such as the FCC under active discussion~\cite{Li:2018jns,FCC:2018evy,FCC:2018vvp,FCC:2018byv}, our analysis focuses on obtaining an updated status on the collider bounds, taking into account recent measurements performed by the ATLAS and CMS experiments~\cite{CMS:2017pet,ATLAS:2017xqs,ATLAS:2018ceg,ATLAS:2022pbd} as reference points. Furthermore, we also comment on potentially improved future electroweak precision constraints and explore novel lepton flavour-violating processes predicted by the type-II seesaw model, which will be measured with higher precision soon. 
\subsubsection{Direct Collider Constraints}
\label{subsec:collider}
\begin{figure}[!t]
	\centering
	\subfigure[]{\includegraphics[width=0.38\linewidth]{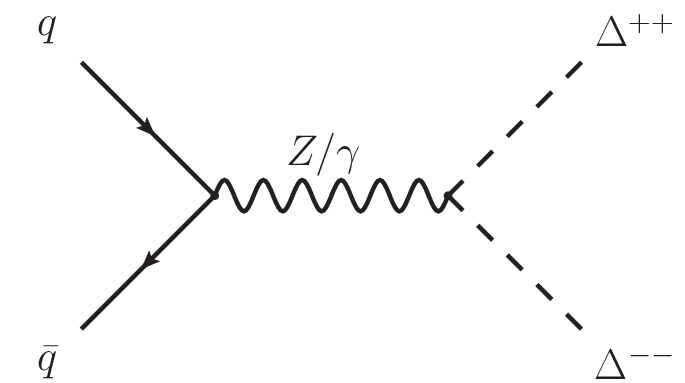}}
	\subfigure[]{\includegraphics[width=0.38\linewidth]{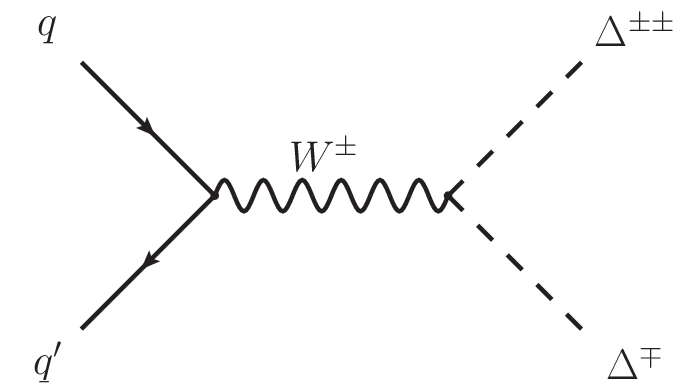}}
	\caption{Representative Feynman diagrams for the production of doubly charged scalars through (a) pair production via neutral-current DY and (b) associated production via charged-current DY processes.\label{fig:proddiags}}
\end{figure}

Firstly, we are going to focus on constraining the parameter space of the type-II seesaw model using the wealth of collider analyses. In particular, we focus on the production of the doubly charged scalar ($\Delta^{\pm\pm}$) which is the primary LHC search channel and the smoking gun signal for the type-II seesaw mechanism. The main processes of interest are the pair production of these doubly charged scalars through the neutral-current Drell-Yan (DY)-like process mediated by virtual $Z/\gamma$ contribution. Another relevant mode of production is via the charged-current interactions mediated by $W$ bosons (see Fig.~\ref{fig:proddiags}). The decay channels of interest for $\Delta^{\pm \pm}$ for these searches are mainly the decays into same-sign lepton pairs ($\Delta^{\pm \pm} \rightarrow l_\alpha^\pm l_\beta^\pm$ with $l = e,\mu$), and the decays into gauge bosons ($\Delta^{\pm \pm} \rightarrow W^\pm W^\pm$). In the limit of $v_\Delta \ll 0.1~\text{MeV} (\ll v)$, the dilepton channel dominates, and therefore, the four-lepton channel ($\Delta^{++} \Delta^{--} \rightarrow l^+ l^+ l^- l^-$) arising from the neutral-current DY process, will provide a clear BSM signature. 
This can be utilised to dramatically suppress expected backgrounds in the SM through sideband analyses of the same-sign mass spectra, which are experimentally under very good control (see~\cite{Williams:2008sh,ATLAS:2016vox,ATLAS:2020rej,ATLAS:2022pbd,ATLAS:2024moy}). The branching ratios related to the exclusive final states of the $\Delta^{\pm\pm}$ decay depends on the neutrino mass-mixing matrix and thus serve as a probe of its structure. 

The ATLAS and CMS collaborations have been able to exclude masses of $\Delta^{\pm \pm}$ lighter than $850~\text{GeV}$ in their analyses, assuming a $100\%$ decay into a light-lepton pair ($ee$, $\mu \mu$, or $e \mu$)~\cite{CMS:2017pet,ATLAS:2017xqs}. From this, model-specific constraints can be obtained by direct rescaling of the branching ratios (which will also include tau contributions). For comparability, we will follow the discussion of~\cite{CMS:2017pet,ATLAS:2017xqs}. For a larger $v_\Delta$, however, the triplet Yukawa couplings become small, and the diboson decay channel dominates (at the price of rising tension with the $\rho$ parameter). The limits set by ATLAS currently are at around $220~\text{GeV}$ for $v_\Delta=0.1~\text{GeV}$~\cite{ATLAS:2018ceg}, which is considerably weaker compared to the constraints from the dilepton channels. 

The collider analyses carried out by ATLAS and CMS are not representative of the full parameter space of the model. They probe specific regions of the triplet vev (in particular, to the very low and very high values of $v_\Delta$ for the two decay channels), whilst assuming mass degeneracy of the scalar triplet masses. These assumptions render the decays of $\Delta^{\pm \pm}$ into same-sign leptons and dibosons phenomenologically dominant. Focussing on the experimentally and theoretically well-motivated four-lepton final states, we will estimate and compare the expected direct future bounds from the HL-LHC phase as well as a future hadron-hadron machine, extrapolating from the current analyses under identical assumptions.

To this end, we employ \textsc{FeynRules}~\cite{Alloul:2013bka} to create a model with normal neutrino mass hierarchy (i.e., $m_{\nu_1} < m_{\nu_2} < m_{\nu_3}$), which is interfaced with \textsc{MadGraph}\_aMC@NLO~\cite{Alwall:2014hca} through the {\sc{Ufo}} package~\cite{Degrande:2011ua,Darme:2023jdn}. Events are generated at center-of-mass energies of $\sqrt{s} = 14~\text{TeV}$ and $\sqrt{s} = 100~\text{TeV}$ for the neutral current DY process: $p p \rightarrow \Delta^{++} \Delta^{--} \rightarrow l^+ l^+ l^- l^-$. This analysis assumes a $100\%$ branching ratio. Our investigation is centred around the search for doubly charged scalars~\cite{Anisha:2021jlz,ATLAS:2017xqs} with parton-level cuts to provide qualitative sensitivities at present and future colliders.
\begin{figure}[!t]
	\centering
	\subfigure[]{\includegraphics[width=0.48\linewidth]{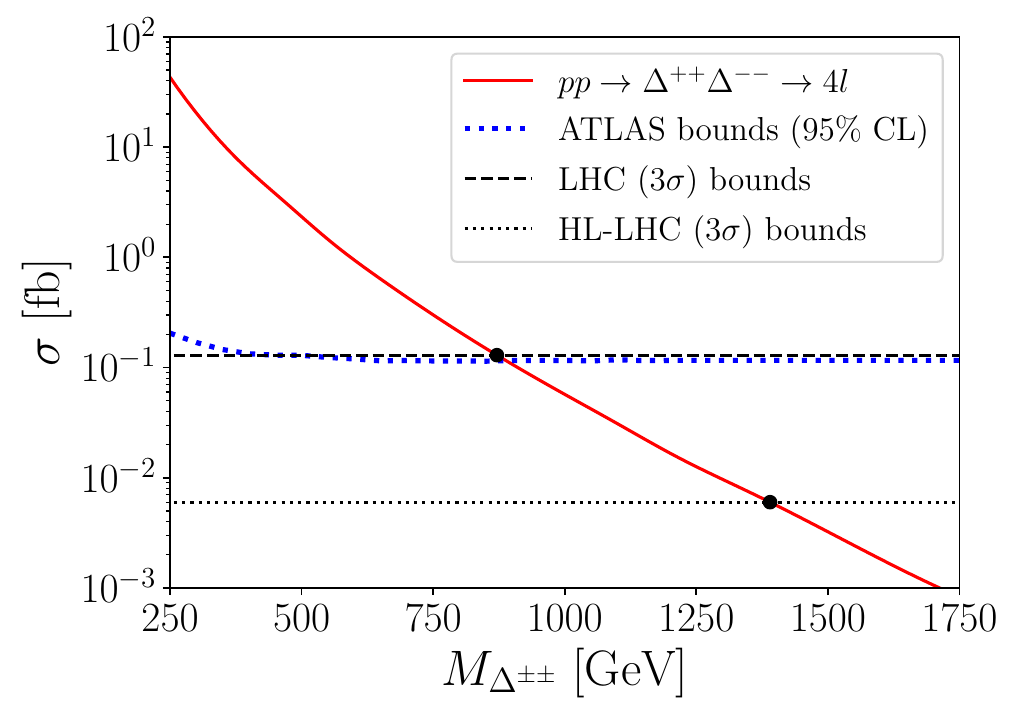}\label{fig:lhc}}
	\subfigure[]{\includegraphics[width=0.48\linewidth]{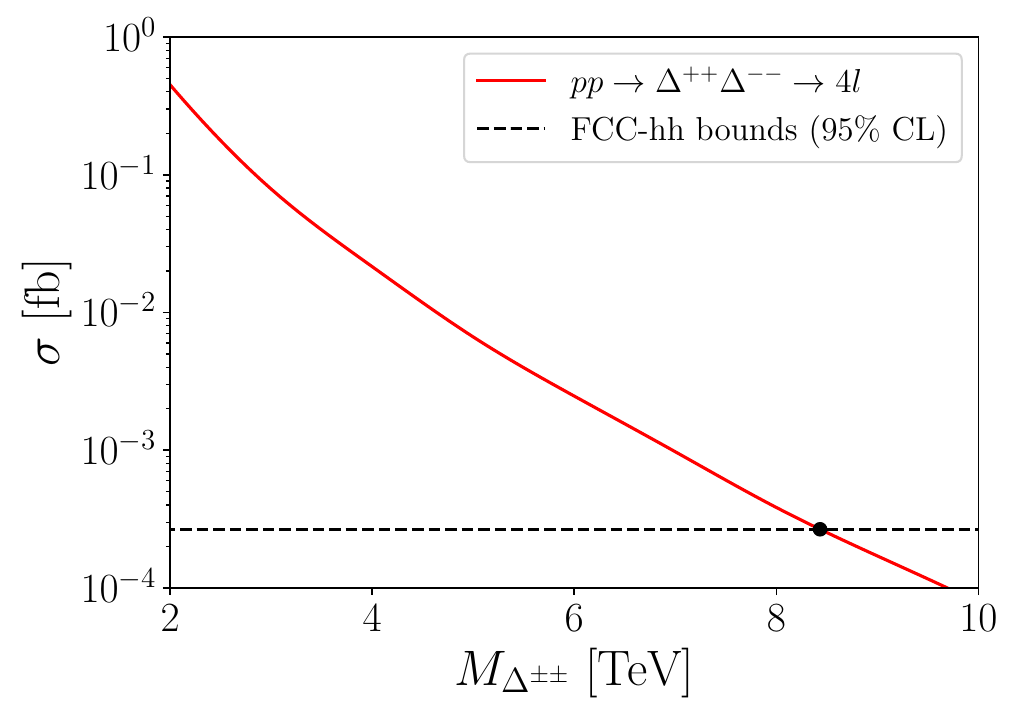}\label{fig:fcc}}
	\caption{Cross-section for pair production of the doubly-charged scalars through neutral DY currents for (a) $\sqrt{s}=14~\text{TeV}$ and (b) $\sqrt{s}=100~\text{TeV}$ in the four-lepton final state (assuming a $100\%$ branching ratio as Ref.~\cite{ATLAS:2017xqs,Fuks:2019clu}). The $3 \sigma$ LHC ($139~\text{fb}^{-1}$) and  HL-LHC ($3~\text{ab}^{-1}$) exclusion bounds are shown by the black dashed and dotted lines respectively on subfigure (a). The $95 \%$ CL on the search for $\Delta^{\pm \pm}$ with a $100 \%$ branching ratio into light leptons, cf. Ref.~\cite{ATLAS:2017xqs}, is represented by the blue dotted line, showing a good agreement of our analysis. The $95\%$ confidence bound for FCC-$hh$ ($30~\text{ab}^{-1}$) is shown by the black dashed line on subfigure (b). These plots show that the LHC is sensitive to doubly charged scalars up to masses $\sim 870~\text{GeV}$ currently, and up to masses $\sim 1390~\text{GeV}$ in its high-luminosity phase. The FCC-$hh$ will be sensitive to similar final states up to a mass scale of $\sim 8.5~\text{TeV}$.\label{fig:colliderbounds}}
\end{figure}
We impose specific cuts, requiring all leptons ($l = e/\mu$) to be within the central part of the detector ($|\eta(l)| < 2.5$) and with a transverse momentum $p_T > 30~\text{GeV}$. Only leptons with no jet activity within the cone radius are considered ($\Delta R (j,l) < 0.4$), and the final states must contain exactly one positively-charged and one negatively-charged lepton pair; otherwise, the event is vetoed. An invariant mass cut is always applied for each lepton pair $M_{l^\pm l^\pm} > 200~\text{GeV}$.

Given that we expect the same-signed leptons to result from $\Delta^{\pm\pm}$ decays, we ensure the consistency of the two masses by calculating,
\newpage
\begin{eqnarray}
\overline{M} = \frac{M_{l^+ l^+} + M_{l^- l^-}}{2}\,,\\
\Delta M = |M_{l^+ l^+} - M_{l^- l^-}|\,.
\end{eqnarray}
The two masses are considered consistent if $\Delta M/ \overline{M} < 0.25$, thus imposing a resonant signal character. Additionally, an event is vetoed if there exists an opposite-signed same-flavor lepton pair with an invariant mass in the range $M_{l^+ l^-} \in [81.2,101.2]~\text{GeV}$ to suppress background resulting from $Z$ decays.

The total number of signal events is then determined as $N_S = \sigma \times {\cal{L}} \times A$, where $A$ represents acceptance. Assuming a good background subtraction which is possible through sideband analyses~\cite{Williams:2008sh,ATLAS:2016vox,Schwartz:2017hep,ATLAS:2020rej,ATLAS:2022pbd}, we calculate the statistical significance at integrated luminosities ${\cal{L}} = 139~\text{fb}^{-1}$ and $3~\text{ab}^{-1}$ at $\sqrt{s} = 14~\text{TeV}$, obtaining $3 \sigma$ mass sensitivities at the LHC and the high-luminosity limit, as depicted in Fig.~\ref{fig:lhc}. We calibrate our acceptance to reproduce the $3 \sigma$ LHC bounds from the ATLAS search for doubly charged scalars reported in~\cite{ATLAS:2017xqs}. The expected $95 \%$ confidence limits (CL) shown by the blue dotted line in Fig.~\ref{fig:lhc} is taken from~\cite{ATLAS:2017xqs} which demonstrates very good agreement as the reference point for our extrapolation. These mass limits also align with those reported in~\cite{ATLAS:2017xqs,CMS:2022cbe}.
Applying the same methodology to $\sqrt{s} = 100~\text{TeV}$, we provide a rough estimate of the mass sensitivity at the FCC-$hh$ with an integrated luminosity of ${\cal{L}} = 30~\text{ab}^{-1}$, as illustrated in Fig.~\ref{fig:fcc}. As usual in searches at low backrgounds, the sensitivity gain when moving to 100 TeV entirely stems from the much larger relevant partonic energy range that can be obtained. The scaling of sensitivities therefore directly reflects the available centre-of-mass energy.

\subsubsection{Electroweak Precision Constraints}
\label{subsec:EWP}
To constrain the model parameter space indirectly (e.g. at a future lepton machine), we focus on constraints from electroweak precision observables, namely the $S$, $T$, and $U$ oblique parameters. Following~\cite{Cheng:2022hbo}, we can obtain the oblique parameters as,
\begin{subequations}
\begin{align}
	\label{eqn:S}
	S &= -\frac{1}{3\pi} \ln{\frac{M_{\Delta^{\pm\pm}}^2}{M_{\Delta^0}^2}}-\frac{2}{\pi}\left[(1-2s_W^2)^2 \xi \left( \frac{M_{\Delta^{\pm\pm}}^2}{M_{Z}^2},  \frac{M_{\Delta^{\pm\pm}}^2}{M_{Z}^2}\right) + s_W^4 \xi \left( \frac{M_{\Delta^{\pm}}^2}{M_{Z}^2}, \frac{M_{\Delta^{\pm}}^2}{M_{Z}^2}\right)\right.\nonumber\\
	&\hspace{2cm}\left.+ \xi \left( \frac{M_{\Delta^{0}}^2}{M_{Z}^2}, \frac{M_{\Delta^{0}}^2}{M_{Z}^2}\right)\right],\\
	T &= \frac{1}{8\pi c_W^2 s_W^2}\left[  \eta \left( \frac{M_{\Delta^{\pm\pm}}^2}{M_{Z}^2}, \frac{M_{\Delta^{\pm}}^2}{M_{Z}^2}\right) + \eta \left( \frac{M_{\Delta^{\pm\pm}}^2}{M_{Z}^2}, \frac{M_{\Delta^{\pm}}^2}{M_{Z}^2}\right)  \right],\\
	U &= \frac{1}{6\pi} \ln{\frac{M_{\Delta^{\pm}}^4}{M_{\Delta^{\pm\pm}}^2 M_{\Delta^0}^2}} + \frac{2}{\pi}\left[(1-2s_W^2)^2 \xi \left( \frac{M_{\Delta^{\pm\pm}}^2}{M_{Z}^2},  \frac{M_{\Delta^{\pm\pm}}^2}{M_{Z}^2}\right) + s_W^4 \xi \left( \frac{M_{\Delta^{\pm}}^2}{M_{Z}^2}, \frac{M_{\Delta^{\pm}}^2}{M_{Z}^2}\right)\right.\nonumber\\
	&\left.+ \xi \left( \frac{M_{\Delta^{0}}^2}{M_{Z}^2}, \frac{M_{\Delta^{0}}^2}{M_{Z}^2}\right)\right] -\frac{2}{\pi} \left[  \xi \left( \frac{M_{\Delta^{\pm\pm}}^2}{M_{W}^2}, \frac{M_{\Delta^{\pm}}^2}{M_{W}^2}\right) + \xi \left( \frac{M_{\Delta^{\pm\pm}}^2}{M_{W}^2}, \frac{M_{\Delta^{\pm}}^2}{M_{W}^2}\right)  \right],
\end{align}
where the functions $\xi$ and $\eta$ are defined as,
\begin{equation}
\begin{split}
	\xi(x,y) &= \frac{4}{9} - \frac{5}{12}(x+y) + \frac{1}{6} (x-y)^2\\ 
	&+ \frac{1}{4}\left[ x^2 -y^2 - \frac{1}{3}(x-y)^3 - \frac{x^2 + y^2}{x-y} \right] \ln{\frac{x}{y}} -\frac{1}{12} d(x,y) f(x,y)\,,\\
	\eta(x,y) &= x+y-\frac{2xy}{x-y}\ln{\frac{x}{y}}\,,\\
	d(x,y) &= -1 + 2(x+y) -(x-y)^2\,, \\
	f(x,y) &= 
	\begin{cases} -2\sqrt{d(x,y)}\left[\arctan{\frac{x-y+1}{\sqrt{d(x,y)}}}-\arctan{\frac{x-y-1}{\sqrt{d(x,y)}}}\right]\,, & d(x,y) > 0 \\ \sqrt{-d(x,y)}\ln{\frac{x+y-1+\sqrt{-d(x,y)}}{x+y-1-\sqrt{-d(x,y)}}}\,, & d(x,y) \leq 0 \end{cases} \,.
\end{split}
\end{equation}
\end{subequations}
Assuming degeneracy of the triplet masses ($M_{\Delta^0} \simeq M_{\Delta^\pm} \simeq M_{\Delta^{\pm\pm}}$), constraints reported by the \textsc{Gfitter} collaboration~\cite{Haller:2018nnx} impose $M_{\Delta^{\pm\pm}} \gtrsim 35~\text{GeV}$ at $95\%$ CL. To get an estimate on the constraints in future from electroweak precision observables~\cite{Fan:2014vta,Gorbahn:2015gxa}, motivated from TLEP~\cite{TLEPDesignStudyWorkingGroup:2013myl} and GigaZ~\cite{Asner:2013psa}, we reduce the uncertainties in $S$, $T$ and $U$ obtained from \textsc{Gfitter} by an order of magnitude, and we obtain a limit on $M_{\Delta^{\pm\pm}} \gtrsim 105~\text{GeV}$ at $95\%$ CL. This implies that even dramatically improved electroweak precision measurements of oblique parameters at future lepton colliders (such as FCC-$ee$) will not lead to significant sensitivity enhancements. Non-oblique vertex corrections, on the other hand, might provide a complementary avenue to probe non-trivial flavour structures in case such high precision related to the electroweak input parameter set of the SM is achieved.

\begin{figure}[!t]
	\centering
	\includegraphics[width=0.24\linewidth]{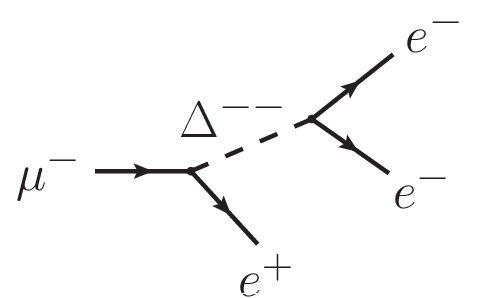}
	\includegraphics[width=0.24\linewidth]{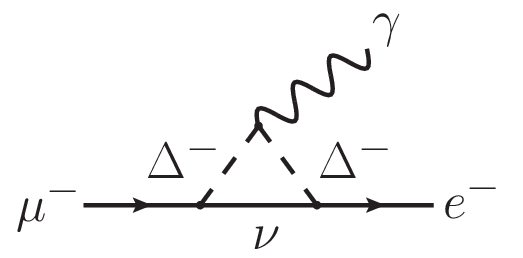}
	\includegraphics[width=0.24\linewidth]{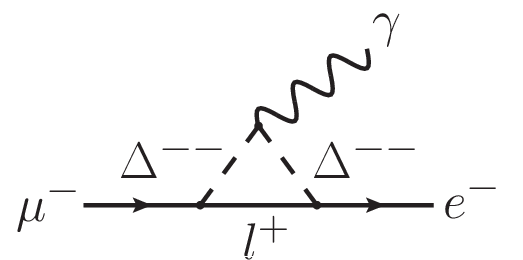}
	\includegraphics[width=0.24\linewidth]{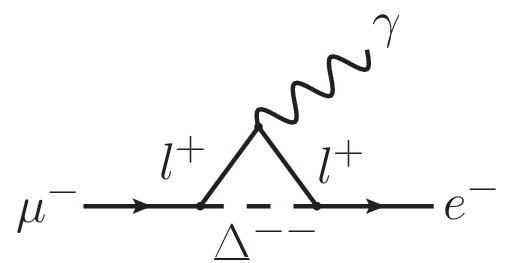}
	\caption{Representative Feynman diagrams for the lepton flavour violating decays $\mu \rightarrow 3 e$ and $\mu \rightarrow e \gamma$.\label{fig:lfvdiags} }
\end{figure}

\subsubsection{Constraints from Lepton Flavour Violating Decays}
\label{subsubsec:lfv}

\begin{figure}[!b]
	\centering
	\subfigure[]{\includegraphics[width=0.49\linewidth]{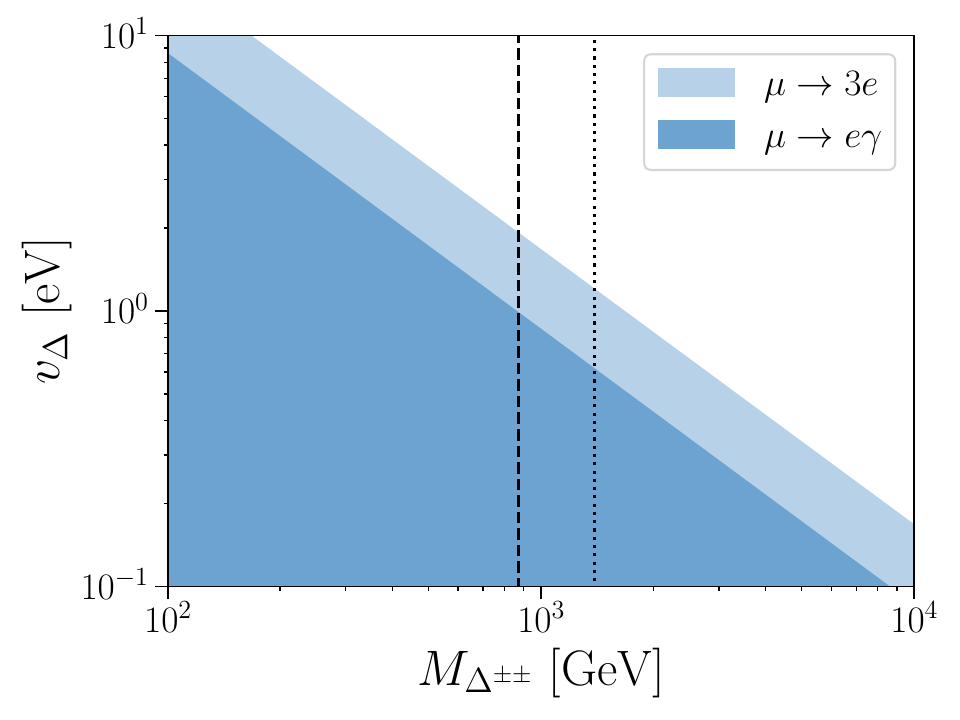}}
	\subfigure[]{\includegraphics[width=0.49\linewidth]{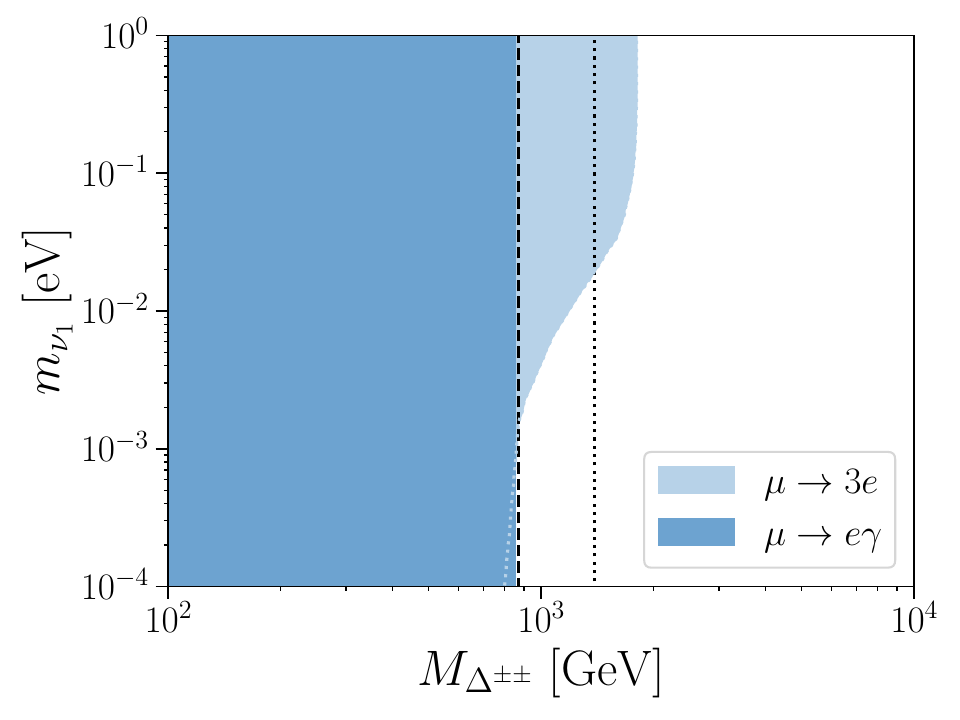}}
	\caption{Constraints from $\mu \rightarrow 3 e$ and $\mu \rightarrow e \gamma$ for (a) $m_{\nu_1} = 0.05~\text{eV}$, and (b) $v_{\Delta} = 1~\text{eV}$. The black dashed and dotted lines on both plots represent the $3\sigma$ LHC and HL-LHC exclusion limits respectively.\label{fig:vev_mnu_mass}}
	\end{figure}

The Yukawa interactions presented in Eq.~\eqref{eqn:yuk}, naturally lead to flavour-changing lepton decays such as $l^-_i \rightarrow l^+_j l^-_k l^+_l$ and $l^-_i \rightarrow l^-_j \gamma$~\cite{Bolton:2022lrg,Akeroyd:2009nu,Dinh:2012bp,Ferreira:2019qpf}. In this work, we are specifically interested in two such lepton flavour violating decays, $\mu \rightarrow 3 e$ and $\mu \rightarrow e \gamma$ (see Fig.~\ref{fig:lfvdiags}). The former arises at tree-level, mediated by the doubly charged scalar ($\Delta^{\pm\pm}$), and its branching ratio is given by,
\begin{equation}
	\label{eqn:mu3e}
	\text{BR}(\mu \rightarrow 3 e) = \frac{|(Y_\Delta)_{ee} (Y_\Delta)_{\mu e}^*|^2}{4 G_F^2 M^4_{\Delta^{\pm\pm}}}\,,
\end{equation}
where $G_F \simeq 1.166 \times 10^{-5}~\text{GeV}^{-2}$ is the Fermi constant. Setting $v_\Delta = 1~\text{eV}$, and $m_{\nu_1} = 0.05~\text{eV}$, and using the global fit of neutrino oscillations data from NuFIT~\cite{deSalas:2017kay}, sets a lower limit on $M_{\Delta^{\pm\pm}} \gtrsim 1650~\text{GeV}$, from the current $\text{BR}(\mu \rightarrow 3 e) < 10^{-12}$ constraint~\cite{SINDRUM:1987nra}, which is well beyond the reach of the current LHC sensitivity regime, as shown in Fig.~\ref{fig:colliderbounds}. $\mu \rightarrow e \gamma$ occurs through one-loop penguin diagrams, mediated by $\Delta^\pm$ or $\Delta^{\pm\pm}$. In the process mediated by $\Delta^\pm$, the photon is emitted exclusively from the $\Delta^\pm$ boson at one-loop, whereas for the $\Delta^{\pm\pm}$ contribution, the photon can be emitted from either $\Delta^{\pm\pm}$ or the charged fermion propagator in the loop. The contributions from both scalars interfere coherently as they couple to leptons with the same chirality, resulting in a branching fraction~\cite{Primulando:2019evb},
\begin{equation}
	\label{eqn:muegamma}
	\text{BR}(\mu \rightarrow e \gamma) = \frac{\alpha_{\text{EM}}}{192 \pi G_F^2} |Y_\Delta^\dagger Y_\Delta|_{\mu e}^2 \left(\frac{1}{M_{\Delta^\pm}^2 }+ \frac{8}{M^2_{\Delta^{\pm\pm}}}\right)^2\,,
\end{equation}
with the electromagnetic fine-structure constant $\alpha_{\text{EM}} \simeq 1/137$. 

Setting $v_\Delta = 1~\text{eV}$, $M_{\Delta^\pm} \sim M_{\Delta^{\pm \pm}}$, and $m_{\nu_1} = 0.05~\text{eV}$ we obtain $M_{\Delta^{\pm\pm}} \gtrsim 850~\text{GeV}$, from the current $\text{BR}(\mu \rightarrow e \gamma) < 3.1 \times 10^{-13}$ constraint~\cite{MEG:2016leq,MEGII:2023ltw}. Assuming degeneracy in the masses of the charged triplet scalars, and using the neutrino oscillation data from NuFIT, both the branching ratios contain three free parameters, $v_\Delta$, $m_{\nu_1}$ and $M_{\Delta^{\pm\pm}}$. The current limits on both these branching ratios can be used to obtain constraints on the resulting parameter space as illustrated in Fig.~\ref{fig:vev_mnu_mass}, where the shaded regions show the parts of the parameter space currently excluded by these constraints. The plots indicate that the regions not excluded by the lepton flavour violating constraints lie mostly beyond the reach of current LHC sensitivity, as well as the HL-LHC, especially for the $\mu \rightarrow 3 e$ process. More precise measurements for these decays are going to push the mass limit even further beyond the reach of colliders. Any potential discovery at the LHC that phenomenologically fits the experimentally clean $\Delta^{\pm\pm}$ expectation could therefore point towards a richer phenomenology of the TeV scale than predicted by the vanilla type-II seesaw model. We turn to this in the next section.

\subsection{Implications of EFT deformations}
\label{sec:eftdef}
In this section, we extend the parameter space of the type-II seesaw model with EFT deformations to analyse the implications of a reduced $\Delta^{\pm\pm}$ mass scale within the LHC's sensitivity reach. 
This provides a new perspective for accommodating or predicing TeV-scale resonances in the light of highly constraining lepton flavour experiments. To this end, we construct the gauge-invariant dimension-6 structures that incorporate at least one $\Delta$ in addition to the usual dimension-6 SMEFT interactions~\cite{Grzadkowski:2010es} and analyse their implications. This construction, which we refer to as a ``BSM-EFT" scenario, is justified by the possibility that the least massive non-SM particle might exist not very far from the electroweak scale, making it accessible to upcoming collider experiments, including improved analysis techniques for the HL-LHC phase~\cite{ZurbanoFernandez:2020cco}.

The potential discovery of charged scalars in future colliders has been extensively studied in the literature within various BSM models, e.g., complex singlets~\cite{Barger:2008jx,Cho:2021itv,Chen:2019ebq,Cho:2023oad,Oikonomou:2024jms}, the two-Higgs doublet model (2HDM)~\cite{Crivellin:2016ihg,Birch-Sykes:2020btk,Anisha:2022hgv,Anisha:2023vvu,Ouazghour:2023plc}, or complex-triplet extensions~\cite{Anisha:2021jlz,Ashanujjaman:2022tdn,Ashanujjaman:2021txz,Padhan:2022hhl,Das:2023zby}. This further motivates the idea of extending these scenarios with EFT interactions to obtain a qualitative understanding of energy scales beyond the reach of current colliders. In parallel, these interactions highlight correlation constraints that are theoretically imposed by the most direct implementation of a certain model.\footnote{For instance, it is known that EFT deformations can significantly modify the phase transition history of the early universe physics, thereby accessing parameter regions that are seemingly excluded by LHC measurements of the 2HDM~\cite{Anisha:2022hgv} with implications for the LHC exotics programme~\cite{Anisha:2023vvu}. Similar observations have been to reconcile the anomalous muonic $g-2$ with leading order constraints of Higgs sector extensions~\cite{Anisha:2021fzf,Anisha:2021jlz}.} The gauge-invariant structures of the relevant operators have been adopted from Ref.~\cite{Banerjee:2020jun,Grzadkowski:2010es}.

\begin{table}[!b]
	\centering
	\renewcommand{\arraystretch}{1.7}
  \small{	\begin{tabular}{|c|c||c|c|}
		\hline 
		$\mathcal{O}_{L\Phi\Delta,ij}^{(1)}$ 
		& $ (\bar{\psi}^c_{L_i} \Delta \psi_{L_j})(\Phi^\dagger \Phi)$ 
		& $\mathcal{O}_{L\Phi\Delta,ij}^{(2)}$ 
		& $ \bar{\psi}^c_{L_i,\alpha} \Delta \Phi^\alpha \Phi^\dagger_{\beta} \psi^\beta_{L_j}$ \\
        $\mathcal{O}_{L\Delta,ij}^{(1)}$
		& $ (\bar{\psi}^c_{L_i} \Delta \psi_{L_j})\text{Tr}[(\Delta^\dagger \Delta)]$ 
		& $\mathcal{O}_{L\Delta,ij}^{(2)}$
		& $ \bar{\psi}^c_{L_i} \Delta \Delta^\dagger \Delta \psi_{L_j}$ \\
		$\mathcal{O}_{ll}^{ijkm}$ 
		& $ (\bar{\psi}_{L_i} \gamma_\mu \psi_{L_j})(\bar{\psi}_{L_k} \gamma^\mu \psi_{L_m})$ 
		& $\mathcal{O}_{ee}^{ijkm}$ 
		& $ (\bar{e}_{i} \gamma_\mu e_{j})(\bar{e}_{k} \gamma^\mu e_{m})$  \\
		$\mathcal{O}_{le}^{ijkm}$ 
		& $ (\bar{\psi}_{L_i} \gamma_\mu \psi_{L_j})(\bar{e}_{k} \gamma^\mu e_{m})$ 
		& $\mathcal{O}^{ij}_{e \phi}$
		& $(\Phi^\dagger \Phi) (\bar{\psi}_{L_i} e_j \Phi)$ \\
		\hline		
\end{tabular}}
\caption{Relevant dimension-6 operators for SM extended by a complex triplet scalar ($\Delta$)~\cite{Banerjee:2020jun,Li:2022ipc,Grzadkowski:2010es}, contributing to $\mu \rightarrow 3 e$, where $i,j,k,m$ are flavour indices, and $\alpha,\beta$ are SU(2) indices. $\Phi$ represents the SM Higgs doublet.} 
\label{tab:operators}
\end{table} 
Given that the constraints from $\mu \rightarrow 3 e$ set the mass scale well beyond the reach of the LHC sensitive regime as discussed in Sec. \ref{subsubsec:lfv}, we focus on the impact of EFT deformations on this particular lepton flavour violating process in our analysis. The gauge-invariant dimension-6 operators for the SM extended by a complex triplet scalar, relevant for $\mu \rightarrow 3 e $ have been tabulated in Tab.~\ref{tab:operators}. We trace their phenomenological relevance again by using \textsc{FeynRules}. We use \textsc{FeynArts}~\cite{Hahn:2000kx}, and \textsc{FormCalc}~\cite{Hahn:1998yk} for the analytical computation of the $\mu \rightarrow 3 e$ branching ratio including the EFT extensions.  

\begin{figure}[!t]
	\centering
	\subfigure[]{\includegraphics[width=0.48\linewidth]{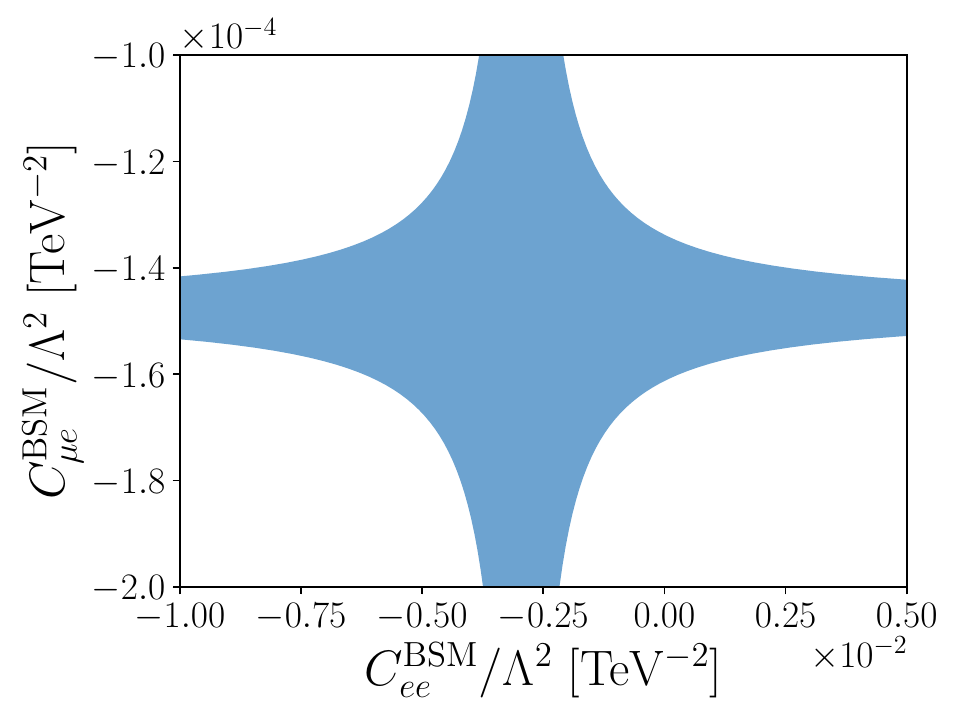}\label{fig:wosmeft}}
	\subfigure[]{\includegraphics[width=0.48\linewidth]{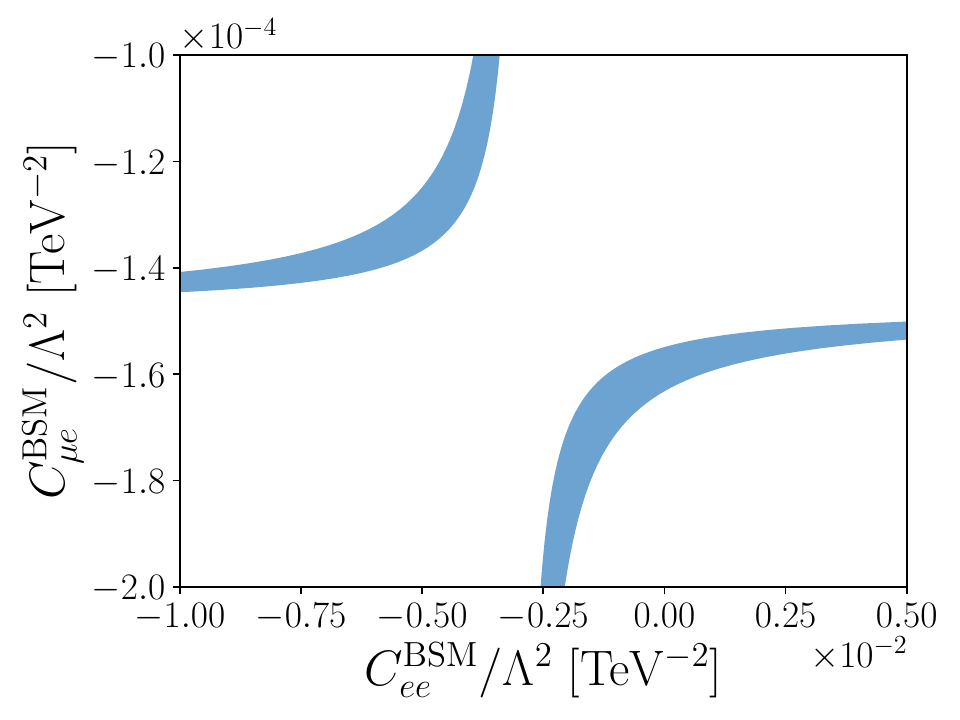}\label{fig:wsmeft}}
	\caption{Allowed regions from $\mu \rightarrow 3 e$ on Wilson coefficients ($C^{\text{BSM}}_{\mu e}$, and $C^{\text{BSM}}_{e e}$) for $M_{\Delta^{\pm\pm}} = 500~\text{GeV}$ (a) without contributions from the SMEFT 4-lepton operators, and (b) setting $C^{1112}_{ll}/\Lambda^2=C^{1112}_{le}/\Lambda^2=C^{1211}_{le}/\Lambda^2=C^{1112}_{ee}/\Lambda^2=2.5\times 10^{-6}~\text{TeV}^{-2}$.\label{fig:BSM-EFT_ops_constraints}}
	\end{figure}

We firstly look at whether the dimension-6 modifications of the Yukawa couplings have an impact on the mass constraints from $\mu \rightarrow 3 e$. The operators $\mathcal{O}_{L\Phi\Delta,ij}^{(1)}$, $\mathcal{O}_{L\Phi\Delta,ij}^{(2)}$, $\mathcal{O}_{L\Delta,ij}^{(1)}$, and $\mathcal{O}_{L\Delta,ij}^{(2)}$ lead to modifications of the Yukawa couplings for $\Delta$ after EWSB. The contributions from $\mathcal{O}_{L\Phi\Delta,ij}^{(1)}$ get absorbed into the Yukawa matrix $Y_{\Delta}$ while generating neutrino masses, and the contributions from $\mathcal{O}_{L\Delta,ij}^{(1)}$ and $\mathcal{O}_{L\Delta,ij}^{(2)}$ are suppressed by $v_\Delta^2 << v^2$ after EWSB. Therefore, the modified Yukawa couplings will have dominant contributions from $\mathcal{O}_{L\Phi\Delta,ij}^{(2)}$ in the parameter region that we have considered above. Upon EWSB, the modified Yukawa couplings (effectively modifying the mass-Yukawa coupling relation of the vanilla type-II seesaw model) can then be written as,
\begin{equation}\label{eq:modified-yukawa}
Y_{ij}^{\text{mod.}} = (Y_\Delta)_{ij} - \frac{C^{\text{BSM}}_{ij} v^2}{2 \Lambda^2}\,,
\end{equation}
where the $C^{\text{BSM}}_{ij}$ is the Wilson coefficient corresponding to the effective operator $\mathcal{O}_{L\Phi\Delta,ij}^{(2)}$. Entering the modified Yukawa couplings into the BR($\mu \rightarrow 3 e$) in Eq.~\eqref{eqn:mu3e}, setting $v_\Delta = 1~\text{eV}$, $m_{\nu_1} = 0.05~\text{eV}$, and $M_{\Delta^{\pm \pm}}=500~\text{GeV}$ such that exotics searches imply a discovery at the LHC, we can constrain the Wilson coefficients to the region depicted in Fig.~\ref{fig:wosmeft}. Since the diagonal Yukawas are $\sim 2$ orders of magnitude larger than the off-diagonal counterparts, Fig.~\ref{fig:wosmeft} clearly illustrates that larger cancellations ($\sim 2$ orders of magnitude) are required in the $ee$ direction compared to $\mu e$.
$\mu \rightarrow 3 e$ also gets direct contributions from SMEFT 4-lepton operators ($\mathcal{O}_{ll}^{ijkm}$, $\mathcal{O}_{ee}^{ijkm}$, and $\mathcal{O}_{le}^{ijkm}$). The contributions from $\mathcal{O}_{e \phi}^{ij}$ however, are suppressed. We therefore do not include these in our analysis. Non-zero Wilson coefficients for the relevant SMEFT operators (in particular $\mathcal{O}_{ll}^{1112}$, $\mathcal{O}_{ee}^{1112}$, $\mathcal{O}_{le}^{1112}$, $\mathcal{O}_{le}^{1211}$) change the allowed parameter space for the Wilson coefficients of $\mathcal{O}_{L\Phi\Delta,ij}^{(2)}$, which can be seen from Fig.~\ref{fig:wsmeft}.

\begin{table}[!b]
	\centering
	\renewcommand{\arraystretch}{1.7}
  \small{	\begin{tabular}{|c|c||c|c|}
		\hline 
		$\mathcal{O}_{eW}$ 
		& $ (\bar{\psi}_{L_i} \sigma^{\mu \nu} e_j)\tau^\alpha \Phi W^\alpha_{\mu \nu}$ 
		& $\mathcal{O}_{eB}$ 
		& $ (\bar{\psi}_{L_i} \sigma^{\mu \nu} e_j) \Phi B_{\mu \nu}$ \\
		\hline		
\end{tabular}}
\caption{Relevant SMEFT dimension-6 operators~\cite{Grzadkowski:2010es}, contributing to $\mu \rightarrow e \gamma$, where $i,j$ are flavour indices, $\mu, \nu$ are Lorentz indices, and $\alpha,\beta$ are SU(2) indices. $\Phi$ represents the SM Higgs doublet.} 
\label{tab:egamma_operators}
\end{table} 
\begin{figure}[!t]
	\centering
	\subfigure[]{\includegraphics[width=0.49\linewidth]{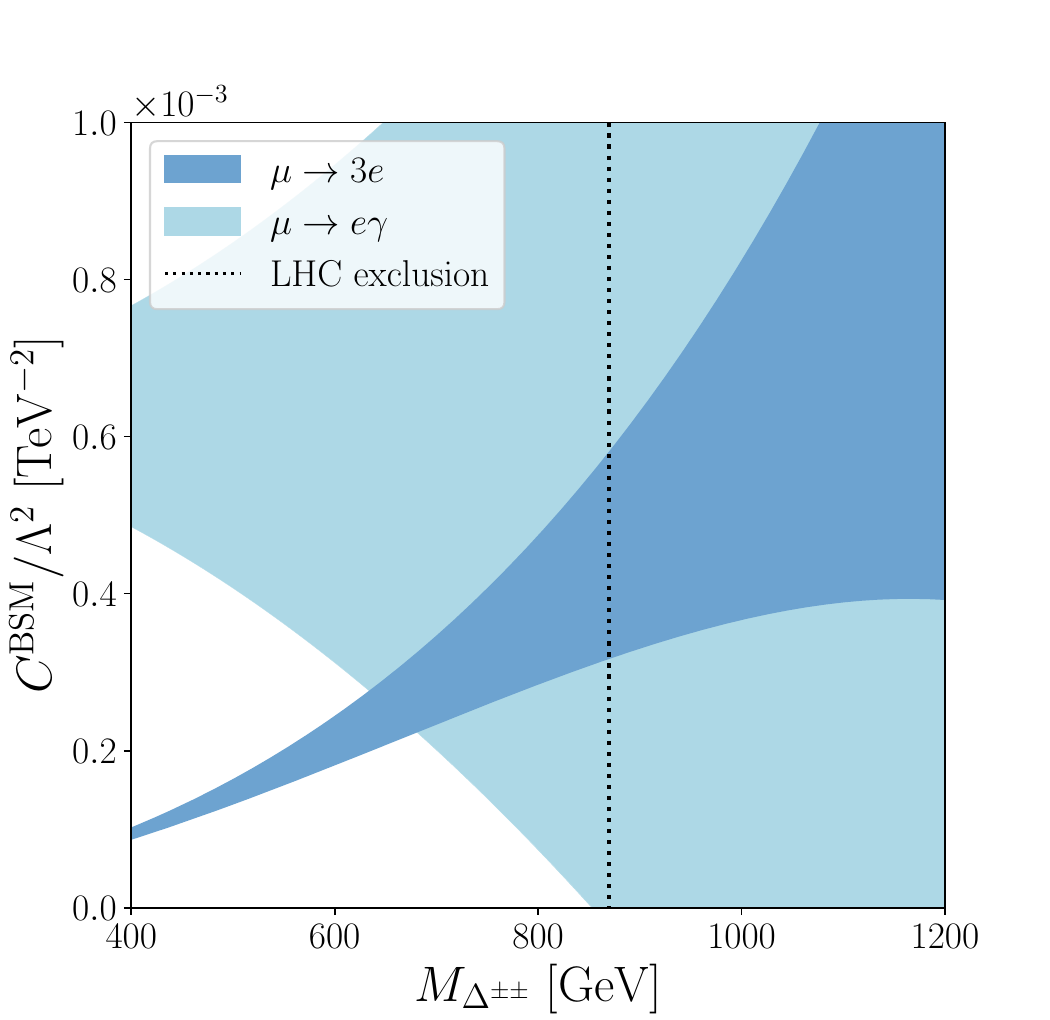}\label{fig:bsm}}
	\subfigure[]{\includegraphics[width=0.49\linewidth]{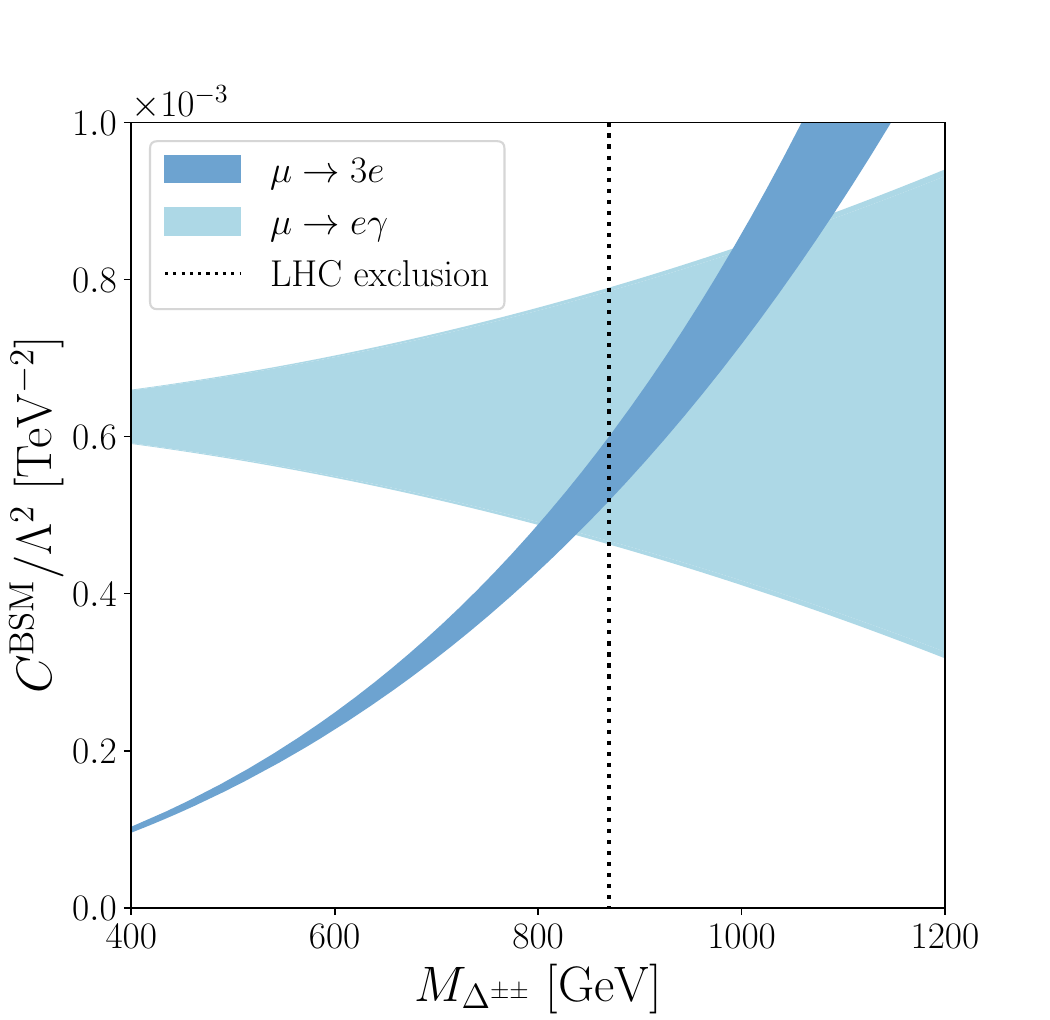}\label{fig:bsm_smeft}}
	\caption{The allowed parameter space of the BSM-EFT Wilson coefficients and $M_{\Delta^{\pm \pm}}$ for (a) $C^{1112}_{ll} = C^{1112}_{le} = C^{1211}_{le} = C^{1112}_{ee} = C_{4f}^{\text{SMEFT}} = 0$, $C_{eB} = C_{eW} = C^{\text{SMEFT}}_{\mu \rightarrow e \gamma} = 0$ and (b) $C^{\text{SMEFT}}_{4f}/\Lambda^2 = 2.5 \times 10^{-6}~\text{TeV}^{-2}$, and $C^{\text{SMEFT}}_{\mu \rightarrow e \gamma}/\Lambda^2 = 0.5 \times 10^{-6}~\text{TeV}^{-2}$. Here, $C^{\text{BSM}}_{e e } = C^{\text{BSM}}_{\mu e } = C^{\text{BSM}}$. The black dotted lines on both plots represent the LHC exclusion limits. \label{fig:BSM-EFT_mass_region}}
\end{figure}
To check how far we can bring the mass scale down, we set $v_\Delta = 1~\text{eV}$, $m_{\nu_1} = 0.05~\text{eV}$ and $C^{\text{BSM}}_{ee} = C^{\text{BSM}}_{\mu e}$. Since we obtain tighter constraints along $\mu e$ compared to $ee$, the latter condition was assumed to obtain the tightest constraints on the BSM-EFT Wilson coefficients, assuming them to be of the same order of magnitude. From the $\mu \rightarrow 3 e$ constraints, we obtain the plots shown in Fig.~\ref{fig:BSM-EFT_mass_region}, which illustrates mass scales well within the region sensitive to the LHC are achievable whilst satisfying constraints from $\mu \rightarrow 3 e$. To present a complete argument on this front, we perform a similar analysis for $\mu \rightarrow e \gamma$. The modified Yukawas outlined in Eq. (\ref{eq:modified-yukawa}) directly contribute to the $\mu \rightarrow e \gamma$ branching ratio given in Eq. (\ref{eqn:muegamma}). Additionally, the SMEFT operators that contribute to $\mu \rightarrow e \gamma$ at tree-level~\cite{Ardu:2021koz} are listed in Tab.~\ref{tab:egamma_operators}. We illustrate our results to this front on Fig.~\ref{fig:BSM-EFT_mass_region} elucidating the fact that we can still probe mass scales within (HL-)LHC sensitivity. As anticipated, cancellations from the SMEFT operators fine-tune and tighten the constraints on the BSM-EFT Wilson coefficients.

\begin{figure}[!t]
	\centering
	\includegraphics[width=0.55\linewidth]{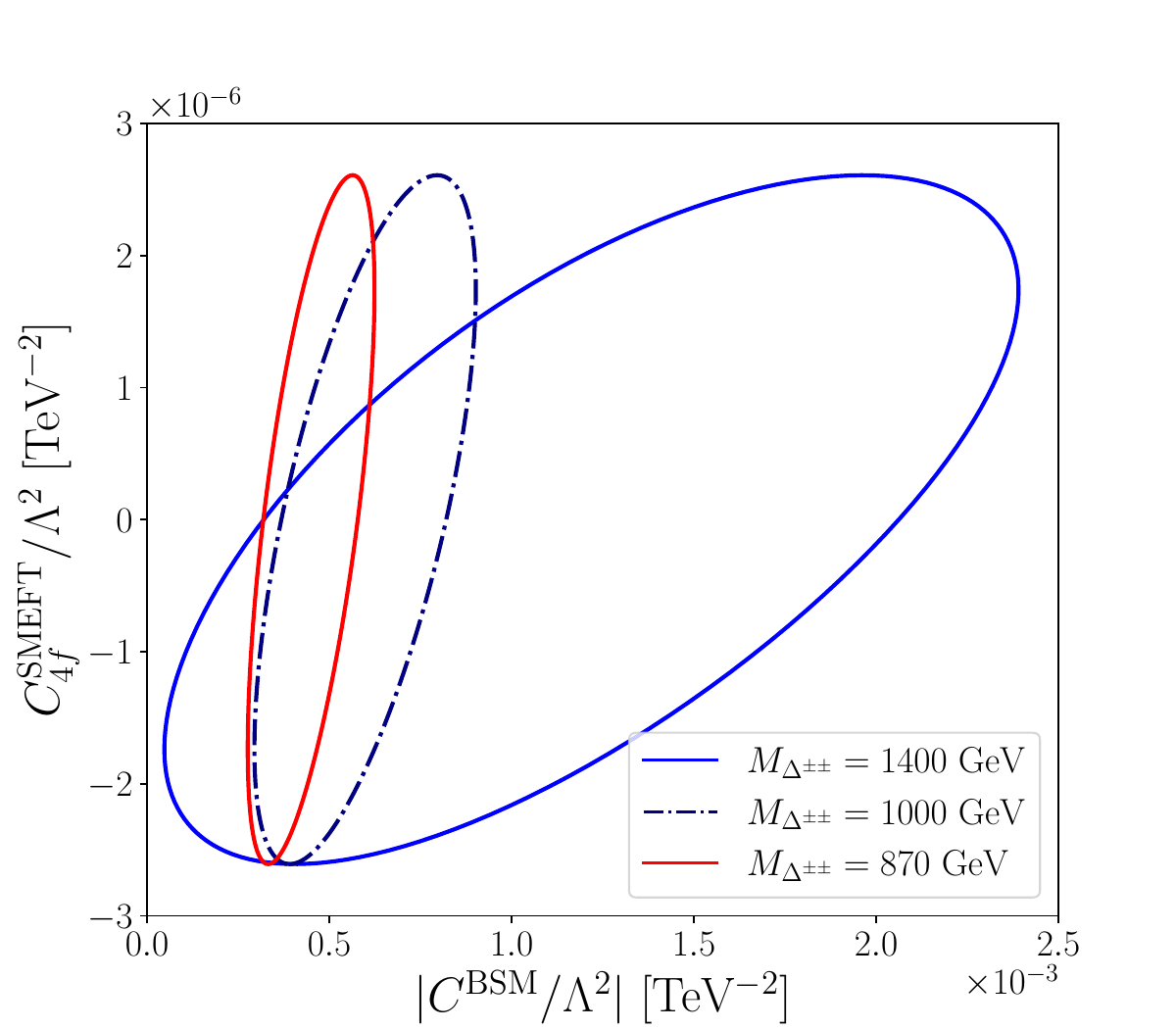}
	\caption{Exclusion contours on SMEFT and BSM-EFT Wilson coefficients for $M_{\Delta^{\pm \pm}} = 870,~1000,~1400~\text{GeV}$ from BR($\mu \rightarrow 3 e$) limits, where $C^{\text{BSM}}_{e e } = C^{\text{BSM}}_{\mu e } = C^{\text{BSM}}$, and $C^{1112}_{ll} = C^{1112}_{le} = C^{1211}_{le} = C^{1112}_{ee} = C_{4f}^{\text{SMEFT}}$.\label{fig:SMEFT_BSM-EFT_region}}
\end{figure}

The combined exclusion contours from $\mu \rightarrow 3 e$ on the Wilson coefficients corresponding to the SMEFT 4-lepton operators and the relevant BSM-EFT operators are presented in Fig.~\ref{fig:SMEFT_BSM-EFT_region}, for three chosen benchmark mass scales, $M_{\Delta^{\pm \pm}} = 870~\text{GeV}$ (representing the current LHC-exclusion limit), $1~\text{TeV}$ and $1.4~\text{TeV}$ (representing the projected HL-LHC exclusion limit). This plot further elucidates the decrease in the allowed parameter space as one tries to bring the mass scale down to the LHC observable region. The constraints on the SMEFT-parametrised Wilson coefficients that improve the tension between the $\Delta^{\pm\pm}$ states and low energy measurements are relatively weak, and further discrimination of the two directions would predominantly be driven by a direct resolution of the cut-off scale, e.g. at a future FCC-hh. It is worth noting that experiments like \textsc{MUonE}~\cite{Abbiendi:2022oks,Kucharczyk:2023dap} are unlikely to provide additional constraints as they predominantly fingerprint the low $Q^2$ behaviour of the scattering process, although the experiment should be sensitive to the signature of $\mu e \to e e $.

\begin{table}[!b]
	\centering
	\renewcommand{\arraystretch}{1.7}
	\small{	\begin{tabular}{|c|c||c|c|}
			\hline 			 
			$\mathcal{O}_{\Phi\Delta D}^{(1)}$&$\big[\Phi^{\dagger}(D_{\mu}\Delta)\big] \big[(D^{\mu}\Delta)^{\dagger}\Phi\big]$  
			&$\mathcal{O}_{\Phi\Delta D}^{(2)}$ 
			&$\big[\Delta^{\dagger}(D_{\mu}\Phi)\big] \big[(D^{\mu}\Phi)^{\dagger}\Delta\big]$ 
			\\
			$\mathcal{O}_{\Phi\Delta D}^{(3)}$&
			Tr$\big[(\Delta^{\dagger}\Delta)\big]\big[(D_{\mu}\Phi)^{\dagger}(D^{\mu}\Phi)\big]$
			&$\mathcal{O}_{\Phi\Delta D}^{(4)}$&
			$(\Phi^{\dagger}\Phi)\;\text{Tr}\big[(D_{\mu}\Delta)^{\dagger}(D^{\mu}\Delta)\big]$
			\\
			\hline			
	\end{tabular}}
	\caption{Dimension-6 operators for SM extended by a complex triplet scalar ($\Delta$)~\cite{Banerjee:2020jun,Li:2022ipc,Grzadkowski:2010es}, relevant for Drell-Yan production of $\Delta^{\pm\pm}$.} 
	\label{tab:operators-direct-detection}
\end{table}

\subsubsection{TeV-modified seesaw at Colliders}
\label{subsec:col_mod_constraints}
The SMEFT four-lepton operators contributing to $\mu \rightarrow 3 e$ can be probed directly at electron-positron machines through the process $e^+ e^- \rightarrow e^\pm \mu^\pm$. We generate events for this process on \textsc{MadGraph}\_aMC@NLO with the \textsc{Ufo} model updated with the relevant dimension-6 operators, for the FCC-$ee$ running at the $Z$ boson pole at $192~\text{ab}^{-1}$, and callibrate our acceptance to reproduce the bounds on 4-lepton operators presented in Refs.~\cite{deBlas:2022ofj,Celada:2024mcf}. The $2\sigma$-bounds on these operators is $|C^{\text{SMEFT}}_{4f}| \lesssim 10^{-4}~\text{TeV}^{-2}$. At a future $ee$-Collider with $\sqrt{s}=3~\text{TeV}$, with an integrated luminosity of $5~\text{ab}^{-1}$ this improves to $|C^{\text{SMEFT}}_{4f}| \lesssim 10^{-5}~\text{TeV}^{-2}$~\cite{deBlas:2022ofj}, therefore implying that $\mu \rightarrow 3 e$ will set more stringent bounds on these Wilson coefficients than future $ee$-Colliders. 

In the context of hadron colliders, these effective operators do not play a role in the production processes of triplet scalars. The dimension-6 BSM-EFT operators that modify the Yukawa couplings do influence the branching ratios; however, given that we have assumed mass configurations with dominant branching ratios for the doubly charged scalar decaying into light-leptons ($e$, $\mu$), this does not imply a sensitivity enhancement compared to our prior analysis. What is more relevant for the production of the doubly-charged scalars is the relevance of the potential EFT deformations to the $Z/\gamma-\Delta^{++}-\Delta^{--}$ vertex. This would induce modifications of the pair production cross sections. 

In Tab.~\ref{tab:operators-direct-detection}, we collect the structures that lead to modifications of the $(Z/\gamma)\Delta^{++}\Delta^{--}$ interactions. It is worth noting that the contributions from these operators would lead to an enhanced (when including squared dimension-6 contributions) cross-section for the DY production of $\Delta^{\pm \pm}$, which in principle should provide updated and improved limits on $M_{\Delta^{\pm \pm}}$ from collider analyses as functions of these Wilson coefficients. The analyses performed in the context of the renormalisable type-II scenario therefore provide conservative estimates of the sensitivity reach, predominantly making the mass scales of the type-II exotic states accessible to collider experiments. More precise measurements of the neutrino oscillation parameters from various experiments, along with updated measurements of the $\rho$-parameter providing a more accurate estimate of $v_\Delta$, which essentially fixes the Yukawa matrix, would lead to a more precise determination of the hadron-collider sensitive region for the model. Similarly, we can expect analyses of the $\Delta^\pm$ states to add additional sensitivity, however, with reduced experimental sensitivity due to a significant amount of missing energy and a smaller electroweak coupling.
\begin{figure}[!t]
	\centering
	\includegraphics[width=0.55\linewidth]{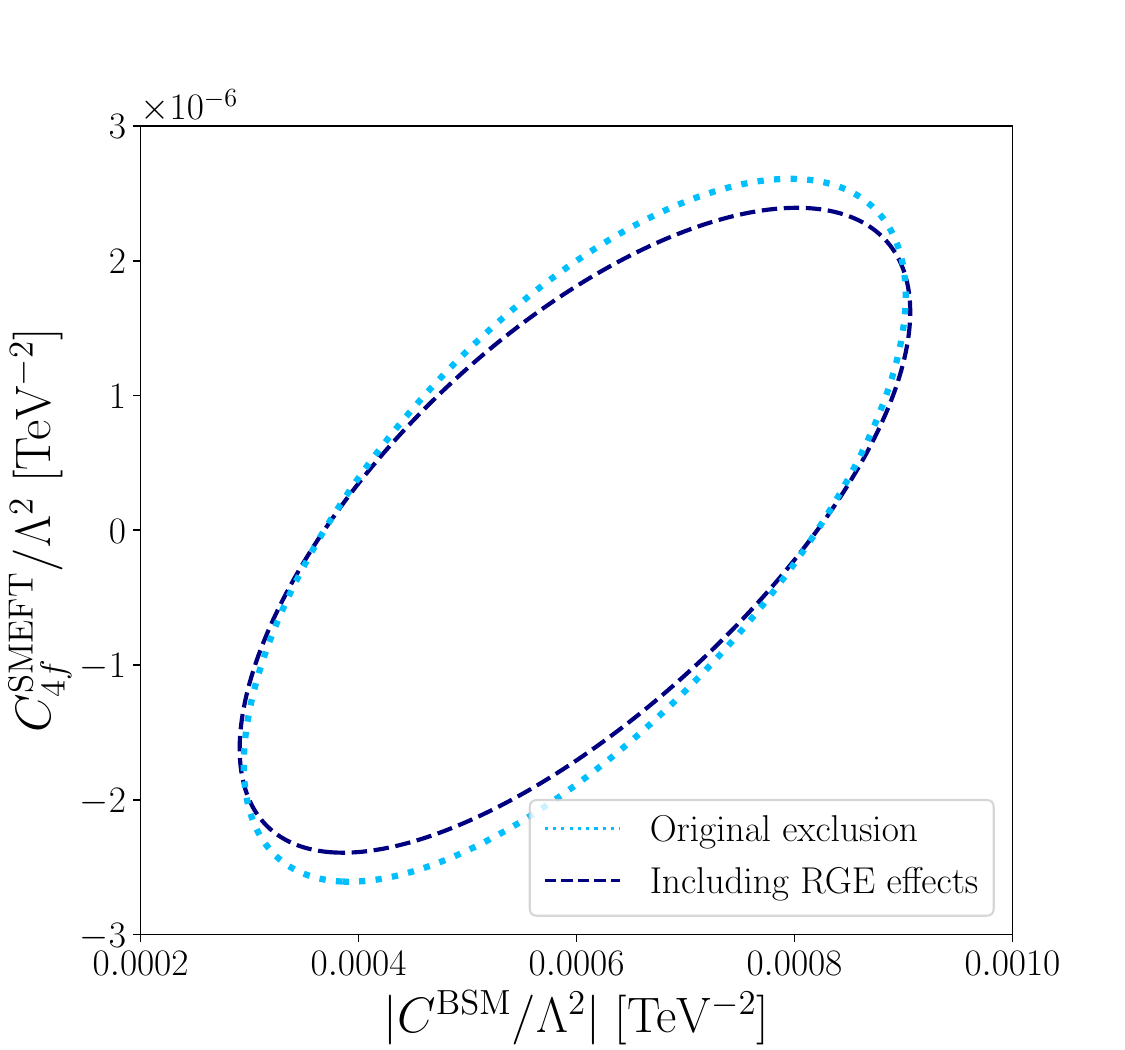}
	\caption{Effect of RGE running on the allowed parameter space of SMEFT and BSM-EFT Wilson coefficients, corresponding to $\mu \rightarrow 3 e$ constraints, for $M_{\Delta^{\pm \pm}} = 1000~\text{GeV}$. Here $C^{\text{BSM}}_{e e } = C^{\text{BSM}}_{\mu e } = C^{\text{BSM}}$, and $C^{1112}_{ll} = C^{1112}_{le} = C^{1211}_{le} = C^{1112}_{ee} = C_{4f}^{\text{SMEFT}}$.\label{fig:SMEFT_BSM-EFT_region_run_2}}
\end{figure}
\subsubsection{Impact of RGE running}
\label{subsec:rge}
As the $\mu$ experiments probe very different energy scales of the theory compared to the collider searches, RGE effects can a priori be important. Below the mass scales of the Higgs sector exotics (which we assume to be degenerate), the running of
the SM couplings are unchanged. To estimate the relevance of these effects, e.g., at the scale of resonance ($M_{\Delta} \gg m_{\mu}$) we can map the modified $\Delta$-contributions on the SMEFT four-fermion interaction that accurately describes $\mu\to 3 e$ at the scale of the muon $m_\mu \ll M_\Delta$. The amplitude is then parametrised by,
\begin{equation}
	C_{ll,\,\text{BSM}}^{1112} (M_{\Delta^{\pm\pm}}^2) = C^{1112}_{ll} (M_{\Delta^{\pm\pm}}^2) - Y_{ee}^{\text{mod.}}(Y_{\mu e}^{\text{mod.}})^{*}  \,,
\end{equation}
where $Y^{\text{mod.}}_{ij}$ can be written in terms of BSM Yukawa couplings $(Y_{\Delta})_{ij}$ and the BSM-EFT Wilson coefficients as shown in Eq.~\eqref{eq:modified-yukawa}. We then compute the RGE flow of the SMEFT Wilson coefficients including the BSM effects, using \texttt{DSixTools}~\cite{Celis:2017hod,Fuentes-Martin:2020zaz} and \texttt{Wilson}~\cite{Aebischer:2018bkb}, between $\Lambda_{\mu}=m_\mu=105~\text{MeV}$ and the scale of the resonance $\Lambda_{\text{UV}}=M_{\Delta}$ to obtain a qualitative estimate of the impact of the renormalisation group flow.

The RG evolution of the Wilson coefficients of the four-fermion operators from $\Lambda_\mu$ up to the scale of resonance $\Lambda_{\text{UV}} = 1~\text{TeV}$ is given by 
\begin{align}
\label{eqn:rge_up}
C_{ll,\,\text{BSM}}^{1112}(\Lambda_{\text{UV}}) &= 1.121 \times C_{ll}^{1112} (\Lambda_\mu) + 0.003 \times C_{le}^{1211} (\Lambda_\mu)\,, \\
C_{ee}^{1112}(\Lambda_{\text{UV}}) &= 1.113 \times C_{ee}^{1112} (\Lambda_\mu) + 0.005 \times C_{le}^{1112} (\Lambda_\mu)\,, \\
C_{le}^{1112}(\Lambda_{\text{UV}}) &= 0.963 \times C_{le}^{1112} (\Lambda_\mu) + 0.021 \times C_{ee}^{1112} (\Lambda_\mu)\,, \\
C_{le}^{1211}(\Lambda_{\text{UV}}) &= 0.968 \times C_{le}^{1211} (\Lambda_\mu) + 0.003 \times C_{ll}^{1112} (\Lambda_\mu)\,.
\label{eqn:rge_up1}
\end{align}
For this specific resonance scale ($M_\Delta = 1~\text{TeV}$), with $v_\Delta = 1~\text{eV}$ and $m_{\nu_1} = 0.05~\text{eV}$ for computing Yukawa couplings, as previously chosen, we substitute the expressions in Eqs.~\eqref{eqn:rge_up}-\eqref{eqn:rge_up1} for the SMEFT Wilson coefficients into the $\mu \rightarrow 3 e$ branching ratio. The resulting exclusions on the parameter space of the Wilson coefficients of the relevant SMEFT and BSM-EFT operators are depicted in Fig.~\ref{fig:SMEFT_BSM-EFT_region_run_2}. We conclude that the RGE flow does not qualitatively change our earlier findings.

\section{Conclusions}
\label{sec:conclusions}
Neutrino masses and oscillations are direct evidence of physics beyond the Standard Model with potentially relevant phenomenological implications for TeV scale physics that is currently explored at the LHC. At the same time, precise measurements of lepton observables corner the parameter space of relevant models at low energies. The quality of these measurements can imply that new physics is far removed from the energy scales that the LHC or even future colliders can explore. What are the circumstances that colliders and potential discoveries at these facilities can still play a pivotal role in fingerprinting the underlying dynamics? 

In this work, we have considered the type-II seesaw mechanism and its effective field theory generalisation with a particular emphasis on the complementarity of low-energy coupling measurements and TeV-scale exotics. Non-minimal versions of this scenario can be described in full generality through effective field theory methods. $\mu\to 3e$ and $\mu\to e\gamma$ when understood as arising from the exchange of the type-II extended scalar sector can push the spectrum to mass scales where sensitivity is difficult to obtain. However, we show that TeV-scale modifications of the type-II scenario can readily bring down mass scales to collider-relevant scales so that future discoveries can be contextualized with low-energy neutrino phenomenology. These effects are correlated with a richer spectrum of interaction above the dynamical degrees of freedom of the type-II scenario, which again can be analysed and constrained at the LHC and future facilities.
As we exploit blind direction in the EFT-extended parameter space, it has to be admitted that the phenomenology discussed in this work is linked to fine-tuning, which we have not addressed dynamically through concrete UV completions. A possibility for the latter might be the extension of ``custodial symmetry'' to multiple Higgs triplets, similar to the ideas presented in, e.g.,~\cite{Dev:2015bta}, which would correlate coupling modifications with the mass-eigenstates of the extended scalar spectrum (similar to how the hierarchy of charged and neutral current interaction strengths is correlated to the gauge boson masses). While we do not attempt to provide a dynamical realisation of this scenario in this work, we stress that a continued search for lepton-flavour relevant new states at the LHC remains a motivated effort, even when high-precision low-energy measurements might suggest the contrary.\\

\medskip
\noindent \textbf{Acknowledgements} ---  We thank Anisha, Dave Sutherland, Graeme Crawford, Joydeep Chakrabortty, and Sabyasachi Chakraborty for helpful discussions, comments and suggestions. C.E. is supported by the UK Science and Technology Facilities Council (STFC) under grant ST/X000605/1 and the Leverhulme Trust under RPG-2021-031. W.N. is funded by a University of Glasgow College of Science and Engineering Scholarship.
	
	
\bibliographystyle{JHEP}
\bibliography{references}
	
\end{document}